\documentclass[aps,prc,twocolumn,groupedaddress,showpacs]{revtex4-1}
\usepackage{graphicx}
\usepackage{comment}
\usepackage{color}

\begin{document}



\title{Hard Substructure of Quenched Jets: a Monte Carlo Study}



\author{K.\,Lapidus$^{\dagger}$ and M.\,H.\,Oliver}

\affiliation{
\mbox{Physics Department, Yale University, 06511~New Haven, CT, USA}\\
\mbox{$^{\dagger}$ corresponding author: kirill.lapidus@yale.edu}
}



\begin{abstract}
Modification of the hard jet substructure in terms of the Soft Drop jet grooming algorithm observables is studied for three different scenarios of jet quenching in a quark-gluon plasma: i) an explicit enhancement of the parton splitting functions, ii) increased soft gluon emissions induced by an in-medium virtuality gain, and iii) energy loss due to a drag force. Despite the fact that first two scenarios both correspond to a radiative energy loss mechanism and lead to similar modifications of parton showers, they are shown to have very different impacts on the momentum balance of hard subjets. Simulations for heavy-ion collisions based on the second scenario are presented and found to be in a good agreement with the experimental data.
\end{abstract}


\maketitle


\section{\label{intro}Introduction}

The suppression of the jet yield in heavy-ion collisions with respect to the reference measured in proton-proton collisions due to the partonic energy loss in the quark-gluon plasma (QGP) is a well-established phenomenon both at RHIC and LHC; see \cite{Connors:2017ptx} for a recent review. 
While the measurement of jet quenching in terms of the nuclear modification factor is essential for inferring the QGP transport coefficients, it contains a limited amount of information. 
Among numerous ingenious approaches to probe the physics of the jet-medium interaction in more detail are studies of jet substructure. Recently, a number of novel jet substructure techniques, initially developed for particle physics applications, were adopted by the heavy-ion community, including the characterization of the hard jet substructure with the Soft Drop (SD) algorithm \cite{Larkoski:2014wba, Larkoski:2015lea}. In this approach, a reconstructed jet is re-clustered with the Cambridge/Aachen (C/A) algorithm \cite{Dokshitzer:1997in, Wobisch:1998wt}, resulting in an angular-ordered tree of jet splittings. The SD algorithm iterates backwards in the C/A clustering history, dropping soft jet branches, until a sufficiently hard splitting is identified. 
At each iteration, the SD algorithm tests the following condition for the splitting formed by the two subjets with transverse momenta $p_{T1}$ and $p_{T2}$:

\begin{equation}
\label{eq:SDcriterium}
z > z_{cut} \left( \frac{\Delta R_{12}}{R} \right)^{\beta},~z = \frac{\mathrm{min}(p_{T1}, p_{T2})}{p_{T1}+p_{T2}},
\end{equation}
where $\Delta R_{12}$ is the angular separation between two subjets, $R$ is the jet radius, $z_{cut}$ and $\beta$ are SD parameters. If condition (\ref{eq:SDcriterium}) is fulfilled, the hard splitting is identified successfully, and the terminal $z$ and $\Delta R_{12}$ values (referred to as $z_{g}$ and $R_{g}$) represent the main SD observables of interest.

In the following we constrain the discussion to the version of the SD algorithm with the parameter $\beta$ set to zero. In that case, for proton-proton collisions, the $z_{g}$ distribution is shown \cite{Larkoski:2015lea, Larkoski:2017bvj} to be linked to the symmetrized parton splitting function:
\begin{equation}
\label{sq:SD}
p\left(z_g\right) = \frac{\bar{P}(z_g)}{\int_{z_{cut}}^{1/2} dz \bar{P}(z)} \Theta(z_g - z_{cut}) + \mathcal{O}\left( \alpha^{2}_s \right),
\end{equation}
where $\bar{P}\left(z \right) = P\left(z \right) + P\left(1 - z \right)$.

In proton-nucleus collisions, preliminary results have been reported by the ALICE collaboration \cite{Caffarri:2017bmh}. The $z_{g}$ distribution has been measured in p-Pb collisions at $\sqrt{s} = 5.02$~TeV for charged jets in the kinematical range $60 < p_{\mathrm T} < 120$~GeV$/c$. The preliminary spectra for minimum-bias collisions were shown to be consistent with the {\sc pythia6} \cite{Sjostrand:2006za} (the Perugia 2011 tunes \cite{Skands:2010ak}) simulations.

In heavy-ion collisions a  measurement of the $z_{g}$ distribution was reported by the CMS Collaboration \cite{Sirunyan:2017bsd} in the kinematical range $140 < p_{\mathrm T} < 500$~GeV$/c$. For the most central ($0-10$\%) Pb+Pb collisions at 5~TeV a significant modification of the $z_{g}$ distribution was observed with respect to the reference spectrum measured in proton-proton collisions. An important aspect of this analysis is a cut on the angular separation of two hard subjets, $\Delta R_{12} > 0.1$, required by the limited spatial resolution of the calorimetry. These results ignited tremendous interest, a number of interpretations has followed \cite{Chien:2016led, Mehtar-Tani:2016aco, Milhano:2017nzm, Chang:2017gkt}, and is growing rapidly. At RHIC, $z_g$ was measured by the STAR Collaboration in pp and Au+Au collisions at 200 GeV \cite{Kauder:2017mhg}. Preliminary results show that the $z_g$ distribution is not modified in contrast to the CMS results.

While in proton-proton collisions the meaning of the Soft Drop grooming/tagging procedure and the analytical behavior of the $z_{g}$ distribution are well understood, the case of heavy-ion collisions and quenched jets is much more complicated. Indeed, a number of theoretical approaches are consistent with the recent measurements, while being based on very different concepts: from medium-modified splitting functions \cite{Chien:2016led} to the jet-induced medium response \cite{Milhano:2017nzm}. This work further explores the behaviour of the SD observables employing different Monte Carlo implementations of jet quenching. The study is based on the publicly available jet-quenching model YaJEM \cite{Renk:2008pp, Renk:2009nz}. YaJEM itself is a modification of the {\sc pythia6} parton shower routine {\sc pyshow} \cite{Bengtsson:1986hr, Bengtsson:1986et, Norrbin:2000uu}, incorporating three scenarios of the in-medium partonic showers. These are: i) FMED, based on the explicit modification (enhancement) of the parton splitting functions, ii) RAD, wherein the jet energy is lost via multiple soft gluon emissions induced by a virtuality gain by the shower partons, and iii) DRAG, wherein momentum and energy loss due to a drag force are implemented.

The paper is organized as follows. Section~\ref{sec:models} introduces three conceptually different implementations of jet quenching and discusses their effect on the SD observables for a sample of jets initiated by a fixed-energy parton propagating in a QGP. Section~\ref{sec:heavy-ion} presents a full simulation of the $z_g$ modification in heavy-ion collisions, based on the most successful configuration of YaJEM, and is followed by Conclusions.

\section{In-medium parton showers and Soft Drop observables \label{sec:models}}

\subsection{Vacuum jets}

\begin{figure}[h]
\includegraphics[width=0.5\textwidth]{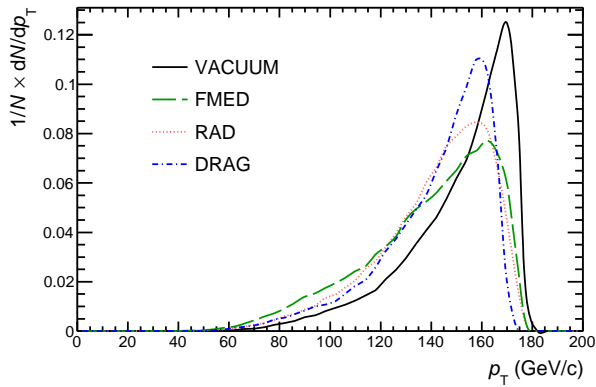}
\caption{\label{fig:jet_spectrum} (Color online) Transverse momentum distributions of leading $R = 0.4$ jets initiated by a 180~GeV gluon in vacuum (solid curve) and for three implementations of jet quenching: FMED (dashed curve), RAD (dotted curve), and DRAG (dash-dotted curve), as described in the text.}
\end{figure}

First, a sample of particle-level vacuum jets initiated by a 180~GeV gluon as simulated with {\sc pythia6} is considered. The $p_{T}$-spectrum of the leading jets reconstructed with the anti-$k_{T}$ jet clustering algorithm with jet radius $R = 0.4$ using FastJet \cite{Cacciari:2011ma} is shown in Fig.~\ref{fig:jet_spectrum} (solid curve).

\begin{figure*}[t]
\includegraphics[width=1.0\textwidth]{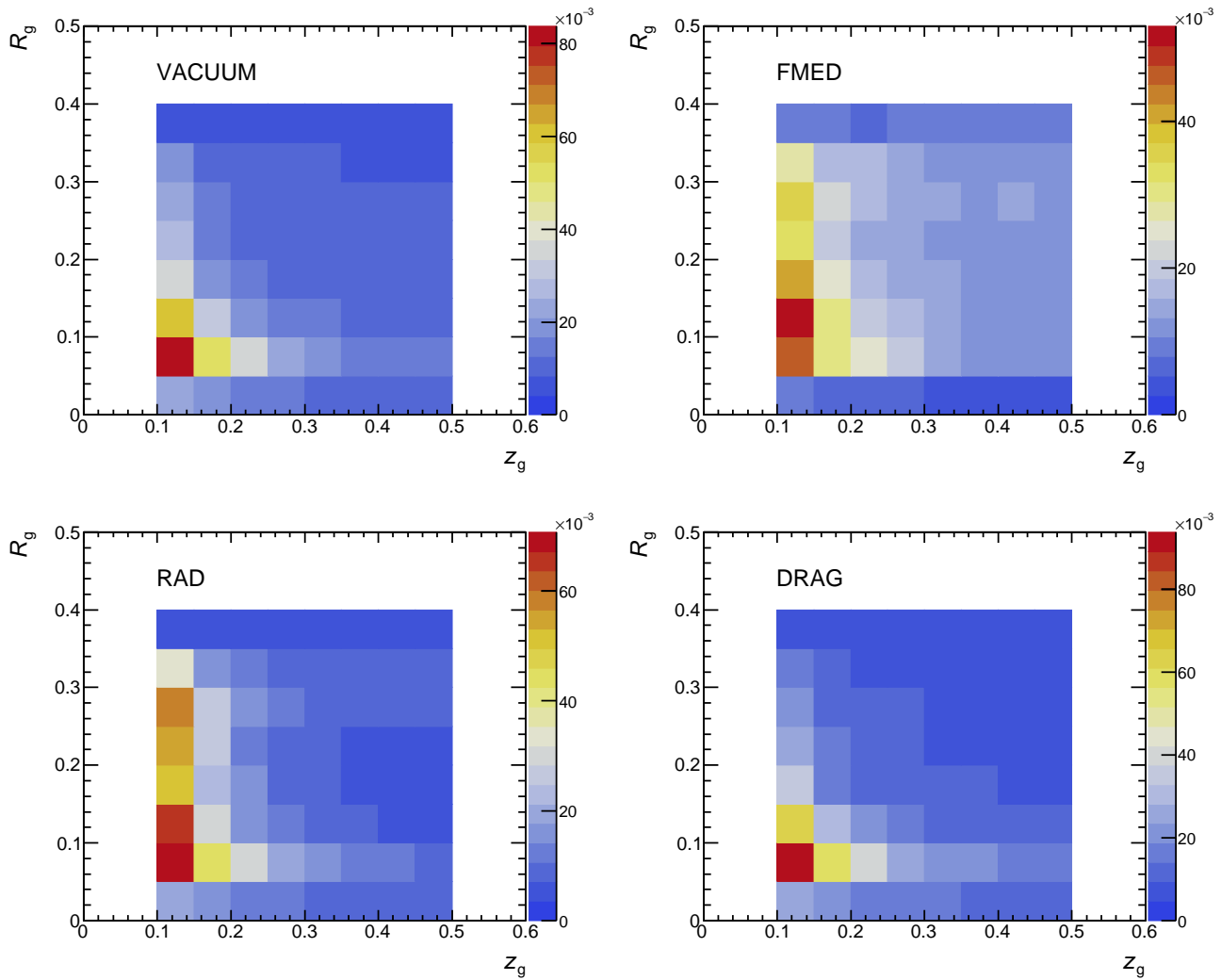}
\caption{\label{fig:zgrg} (Color online) $(z_g, R_g)$ distributions of leading jets initiated by 180~GeV gluons in vacuum (top left) and for three different implementations of the jet quenching: FMED (top right), RAD (bottom left), and DRAG (bottom right).}
\end{figure*}

Figure~\ref{fig:zgrg} (left, top) shows the two-dimensional distribution of the hard splittings, identified by the SD algorithm ($\beta = 0$, $z_{cut} = 0.1$), in the ($z_{g}$, $R_{g}$) plane. As expected \cite{Larkoski:2015lea} and measured in proton-proton collisions \cite{Larkoski:2017bvj}, the distribution is peaked at low $z_{g}$ and $R_{g}$ values, reflecting the corresponding singularities of quantum chromodynamics.

A remark on the difference between quark and gluon jets in the context of the SD observables is necessary. While the $z_g$ distribution is approximately the same for quark- and gluon-initiated jets, the $R_{g}$-distributions are very different. Gluon jets, being broader, are characterized by a much broader $R_{g}$ distribution.
For example, given the sample of leading vacuum jets initiated by partons with a fixed momentum of 180~GeV, while $\approx63\%$ of gluon-initiated SD-tagged jets pass the requirement of $\Delta R_{12} > 0.1$, only $\approx33\%$ of quark jets survive this cut. It can be expected, therefore, that also in heavy-ion collisions, any chosen $\Delta R_{12}$ cut enhances the fraction of gluon jets. The exact strength of this additional bias is controlled by the in-medium modification of the quark/gluon $R_{g}$-distributions.

In the following we will assess the effect on the Soft Drop observables, in particular the ($z_{g}$, $R_{g}$) landscape, of the different implementations of the jet quenching for a stand-alone setup: jets initiated by gluons with a fixed initial energy of 180 GeV, which traverse a fixed trajectory in the expanding QGP environment. The resulting transverse momentum spectra of quenched jets as simulated with three different models are shown in Fig.~\ref{fig:jet_spectrum}. For each model, the strength of the quenching is deliberately chosen to ensure comparable average jet energy losses.

Similar comparative studies of different in-medium parton shower implementations for conventional jet quenching observables and selected jet shapes were performed in \cite{Renk:2009nz} and \cite{Renk:2009hv}, respectively. 

\subsection{FMED}
In the FMED scenario, the parton splitting functions are modified by enhancing their singular parts following the approach proposed in \cite{Borghini:2005em}.
For example, the gluon splitting function ($g \to gg$) becomes
\begin{equation}
\label{eq:BW}
P_{gg}(z) = 2 C_{A}\left( \frac{1+f_{med}}{z} + \frac{1+f_{med}}{1-z} +z\left(1-z\right) -2 \right),  
\end{equation}
where a free parameter $f_{med}$ controls the strength of the modification, $f_{med} = 0$ restores the vacuum splittings.

\begin{figure*}[t]
\includegraphics[width=1.0\textwidth]{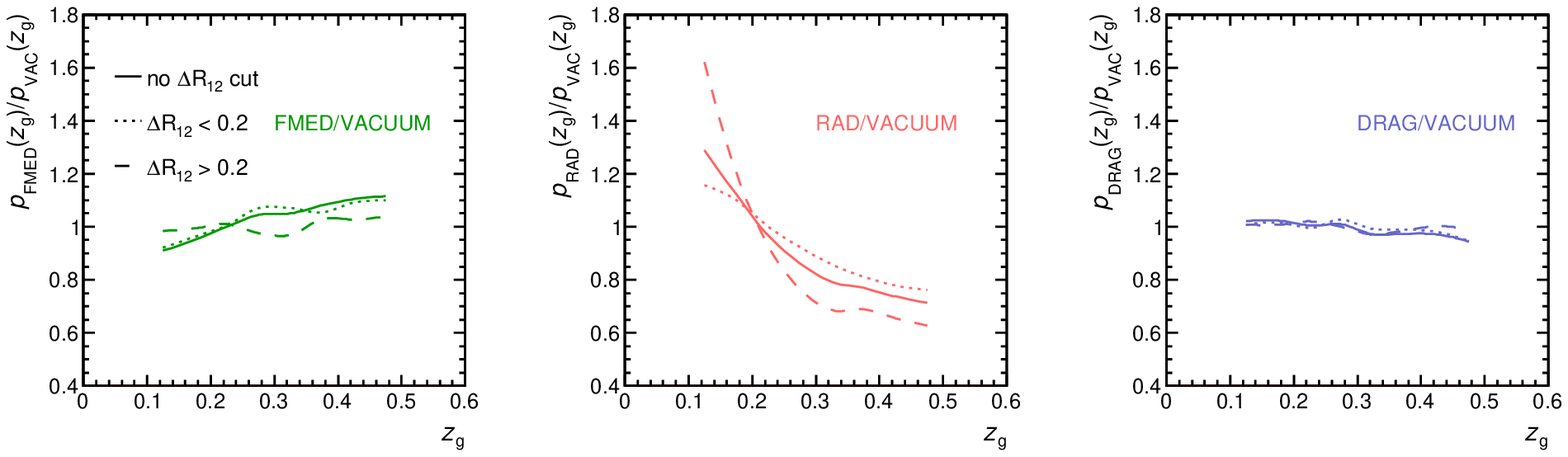}
\caption{\label{fig:zg_ratio} (Color online) Ratios of the $z_g$ distributions of leading jets initiated by a 180~GeV gluon in vacuum and for three different implementations of jet quenching: FMED/vacuum (left), RAD/vacuum (center), and DRAG/vacuum (right). Solid curves -- no $\Delta R_{12}$ cut, dashed curves -- $\Delta R_{12} > 0.2$, dotted curves -- $\Delta R_{12} < 0.2$.}
\end{figure*}

\begin{figure}[h]
\includegraphics[width=0.5\textwidth]{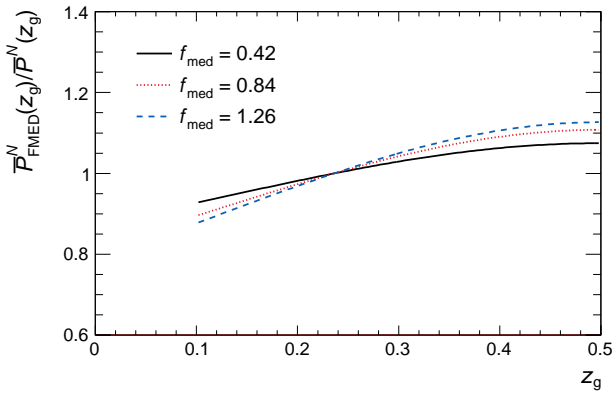}
\caption{\label{fig:Pbar_BW} Ratios (FMED/vacuum) of the symmetrized and normalized $g \to gg$ splitting functions for different values of $f_{med}$ in Eq.~(\ref{eq:BW}). Here, $\bar{P}^{N}\left(z_{g}\right) = \bar{P}\left(z_{g}\right) / \int_{z_{\mathrm{cut}}=0.1}^{0.5} \bar{P}\left(z\right)dz$.}
\end{figure}

For the quenched jet sample the parameter $f_{med}$ is set to 0.42. The resulting distribution of the SD-identified hard splittings in the ($z_{g}$, $R_{g}$) plane is shown in Fig.~\ref{fig:zgrg}, top right. A significant increase of large $R_g$ configurations is observed: $\approx78\%$ of tagged jets are located in the region $R_{g} > 0.1$. This is a direct consequence of the parton shower broadening due to the increased splitting probability. The ratio of the $z_{g}$ distributions (FMED/vacuum) is plotted in Fig.~\ref{fig:zg_ratio}, left panel. A slight increase of the ratio at $z_g \sim 0.5$ is observed. This effect is partially driven by a small change in the shape of the medium-modified splitting function as illustrated in Fig.~\ref{fig:Pbar_BW}, where the ratios of normalized and symmetrized splitting functions (FMED/vacuum) are shown for different values of $f_{med}$. Given the overall small magnitude of the $z_{g}$ modification in the FMED scenario, no significant dependence on the $\Delta R_{12}$ cut is observed (cf. dashed and dotted curves in Fig.~\ref{fig:zg_ratio}, left panel). 

\subsection{RAD}
\label{sec:yajemd}

In the RAD scenario each parton of the shower acquires additional virtuality while propagating in the QGP (cf. \cite{Renk:2009nz} for the details of the implementation). The transport coefficient $\hat q$ quantifies the virtuality gain per pathlength and is assumed to be related to the local energy density $\epsilon$:
\begin{equation}
\label{eq:yajemd}
\hat q(\xi) = \kappa \times \epsilon^{3/4}(\xi),
\end{equation}
where $\kappa$ is a free parameter regulating the parton-medium coupling and $\xi$ is the space-time coordinate. The splitting functions are not modified in this scenario. Instead, the gain in the virtuality affects the kinematics of the shower and opens the phase space for copious soft gluon emissions, resulting, very similar to the previously considered FMED scenario, in a softening and broadening of the partonic shower.

The distribution of hard splittings in the ($z_{g}, R_{g}$) plane for the RAD sample is shown in the left bottom panel of Fig.~\ref{fig:zgrg}. Analogous to the FMED scenario, the broadening of the parton shower is reflected in an increase of the large-$R_{g}$ configurations ($\approx70\%$ of the entries satisfy the $R_{g} > 0.1$ condition). The ratio of the $z_{g}$ distributions (RAD/vacuum) is shown in Fig.~\ref{fig:zg_ratio}, central panel. This implementation of jet quenching leads to an enhancement of the SD-identified hard splittings located at low-$z_{g}$ ($z_{g} \approx 0.1-0.2$) values, reflecting the angular and momentum structure of medium-induced soft emissions. In comparison to FMED, in the RAD scenario the softer branch of the hard splitting is characterized on average by a larger number of constituents and their softer transverse momentum distribution. These two features in combination lead to a pronounced enhancement of the low-$z_{g}$ configurations. Within the RAD scenario the strength of the $z_{g}$ modification distinctly depends on the $\Delta R_{12}$ selection, and the innermost splittings of the jet are less affected. Indeed, a $\Delta R_{12} > 0.2$ ($\Delta R_{12} < 0.2$) cut leads to an increased (decreased) magnitude of the $z_{g}$ modification as seen in Fig.~\ref{fig:zg_ratio}, central panel.

\subsection{DRAG}
\label{sec:yajemdrag}

The most trivial case in terms of the $z_{g}$ modification is the drag force scenario, in which the momentum (and energy) of each parton in the shower is reduced according to:
\begin{equation}
\label{eq:yajeme}
\hat q_{E} = \kappa_{E} \times \epsilon^{3/4}(\xi).
\end{equation}

As for the previously discussed models (FMED and RAD), the consequences of the DRAG approach for the SD observables are shown in Fig.~\ref{fig:zgrg} and Fig.~\ref{fig:zg_ratio}. The hard substructure of jets quenched within the drag force model is essentially unmodified, only a minor suppression of the large-$z_{g}$ configurations is observed.

\begin{figure*}[t]
\includegraphics[width=1.0\textwidth]{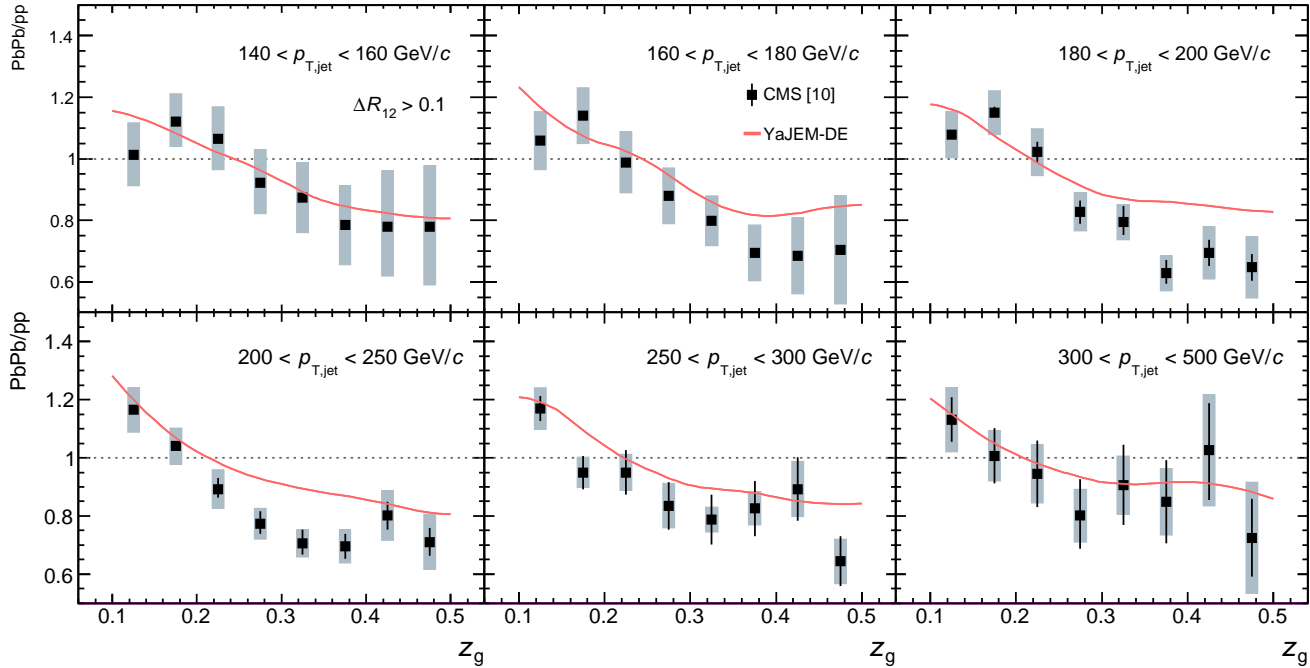}
\caption{\label{fig:cms_drcut} (Color online) Ratios of the $z_g$ distributions in PbPb and pp collisions at 5 TeV: markers are experimental data \cite{Sirunyan:2017bsd}, curves represent YaJEM-DE simulations. }
\end{figure*}

\section{$Z_g$ modification in heavy-ion collisions: YaJEM-DE\label{sec:heavy-ion}}

So far, three scenarios of jet quenching as implemented in the YaJEM code have been introduced. In practice, YaJEM allows one to perform simulations within any combination of these scenarios. The most successful version of the code is YaJEM-DE, introduced in \cite{Renk:2011qf} and tested against a large body of jet-quenching observables, e.g. \cite{Renk:2011aa, Renk:2012cb, Renk:2012hz}. This model is based on the RAD scenario (soft gluon emissions due to the in-medium virtuality gain, cf.~\ref{sec:yajemd}) with a small ($\sim 10\%$) contribution of elastic energy loss (as constrained by the di-hadron correlations measurement \cite{Renk:2011qf}), modeled with the DRAG model (cf.~\ref{sec:yajemdrag}).

Here we outline our implementation of the simulation workflow with YaJEM. The first prerequisite for the simulations is the space-time evolution of the QGP energy density, necessary for the treatment of the parton propagation, according to Eq.~\ref{eq:yajemd}. For this study, we employ a Bjorken-like longitudinal expansion model to simulate the evolution of the QGP created in Pb-Pb collisions at 5~TeV. 
In the next step, YaJEM-DE simulations are performed for all possible trajectories of hard-scattered partons. Finally, in the analysis stage, a weighting procedure is applied that takes into account: i) the transverse distribution of the hard-scattering vertices according to the overlap function of two nuclei, and ii) the quark/gluon parton fractions and energy distributions. The only free parameter in the YaJEM-DE simulations is the coefficient $\kappa$ in Eq.~\ref{eq:yajemd} (as the ratio $\kappa/\kappa_{E}$ is fixed to $0.8/0.1$ \cite{Renk:2011qf}). It is chosen such that the simulations reproduce the charged-hadron nuclear modification factor $R_{AA}\left(p_{T}\right)$ measured in central Pb-Pb collisions at LHC energies. Finally, jets are reconstructed with the anti-$k_{T}$ algorithm with $R=0.4$ and the Soft Drop observables are calculated.

The ratio of the $z_{g}$ distributions reconstructed in Pb--Pb collisions to the pp distributions as simulated with {\sc YaJEM-DE} is shown in Fig.~\ref{fig:cms_drcut}. A $\Delta R_{12} > 0.1$ cut is applied for the comparison with the CMS results. Similar to the stand-alone simulations within the RAD scenario (cf. \ref{sec:yajemd}), the ratios of the $z_{g}$ distributions show a slight enhancement in the low-$z_{g}$ region and suppression at large $z_{g}$ values. Apart from one $p_{T}$-bin ($200 < p_{\mathrm{T, jet}} < 250$~GeV$/c$), YaJEM-DE simulations describe the experimental data very well. For the discussed simulations, the $\Delta R_{12} > 0.1$ cut has a major effect. Indeed, simulations without this cut, i.e. accepting all $R_{g}$ configurations, reveal only a very weak modification in the $z_g$ ratios, resulting in a poor agreement with the data, whereas an increase of the angular separation cut to a larger value, $\Delta R_{12} > 0.2$, leads to a slightly increased modification, following the observations drawn from the stand-alone RAD simulations (cf. Fig.~\ref{fig:zg_ratio}, central panel). 

\section{Conclusions \label{sec:summary} }

This work explored how three different concepts of the in-medium parton shower modification--all implemented within the YaJEM code--are reflected in the Soft Drop observables, in particular in the $(z_g, R_g)$ plane, where each considered implementation of jet quenching is shown to leave a unique footprint.

Two different implementations of radiative energy loss, FMED and RAD, lead to an increase of the large-$R_{g}$ configurations; this is a general feature of any jet-quenching model that broadens parton showers. The impact on the $z_{g}$ distribution is, however, specific for each of the scenarios. While for the FMED scenario the distribution shows an increase at large $z_{g}$ values, an opposite trend is observed for the RAD model. Simulations for heavy-ion collisions, based on the RAD scenario, are shown to be consistent with the recent CMS measurement.
An almost negligible modification is observed for the drag force scenario DRAG, which approximately preserves the hard substructure of a jet, at least in the considered kinematical range.

The quantitative characteristics of the soft radiation (or, more generally, of any other substructure-altering mechanism) are imprinted in the ($z_{g}$, $R_{g}$) landscape of the Soft Drop tagged jets. While a complete triply-differential measurement of groomed jets ($p_{T}$, $z_{g}$, $R_{g}$) is a formidable experimental undertaking, such measurement for 2-3 broad $R_{g}$ bins is feasible and essential to constrain the swiftly growing number of analytical and Monte Carlo calculations that are largely compatible with the experimental data.  

This work is restricted to the exploration of the novel class of the Soft Drop observables. Clearly, a larger set of jet substructure observables would further challenge the models of the in-medium jet modification. In fact, one particularly important constraint is imposed by a recent jet mass measurement in heavy-ion collisions \cite{Acharya:2017goa}. While the class of jet quenching models with the enhanced soft emissions, such as YaJEM-DE, is comparable with the $z_{g}$ measurement by the CMS collaboration, such models by construction tend to predict an increase of the jet mass in medium, contrary to the experimental findings.

Both experimental measurements and theoretical calculations concentrated so far on the simplest version of the Soft Drop algorithm, namely setting the $\beta$ parameter in Eq.~\ref{eq:SDcriterium} to zero. Performance of the $\beta \neq 0$ setting for the case of quenched jets, and potential benefits of it in terms of the sensitivity to a particular aspect of jet quenching, are yet to be explored.

\begin{acknowledgments}
This work was supported by the U.S. Department of Energy under grant number DE-SC0004168. K.L. is thankful to Prof. Joern Putschke for inspiring ideas on the jet substructure observables in the context of relativistic heavy-ion collisions, and to Prof.~Helen Caines for numerous valuable discussions.
\end{acknowledgments}

\bibliography{zg_pheno}

\end{document}